\documentclass[final,5p,times,twocolumn]{elsarticle}

\usepackage{multirow,booktabs}

\usepackage{amsmath}

\usepackage{listings}
\usepackage{float}
\usepackage{rccol}
\usepackage[table]{xcolor}
\usepackage{algorithm2e}
\usepackage{wrapfig}

\usepackage{amssymb}
 \usepackage{makecell}

\journal{NeuroImage}

\begin{document}
\begin{frontmatter}

\title{Explicitly Linking Regional Activation and Function Connectivity: Community Structure of Weighted Networks with Continuous Annotation}

\author[label1,label2]{Andrew C. Murphy}
\author[label1,label2]{Shi Gu}
\author[label1]{Ankit N. Khambhati}
\author[label3]{Nicholas F. Wymbs}
\author[label3]{Scott T. Grafton}
\author[label1,label2]{Theodore D. Satterthwaite}
\author[label1,label4]{Danielle S. Bassett  \fnref{label0}}
  \fntext[label0]{\scriptsize To whom correspondence should be addressed: dsb@seas.upenn.edu}
\address[label1]{Department of Bioengineering, University of Pennsylvania, Philadelphia, PA 19104, USA}
\address[label2]{Perelman School of Medicine, University of Pennsylvania, Philadelphia, PA 19104, USA}
\address[label3]{Department of Psychological and Brain Sciences, University of California Santa Barbara, Santa Barbara, CA 93106, USA}
\address[label4]{Department of Electrical \& Systems Engineering, University of Pennsylvania, Philadelphia, PA 19104, USA}

\begin{abstract}
{A major challenge in neuroimaging is understanding the mapping of neurophysiological dynamics onto cognitive functions. Traditionally, these maps have been constructed by examining changes in the activity magnitude of regions related to task performance. Recently, the emerging field of network neuroscience has produced methods to map connectivity patterns among many regions to certain cognitive functions by drawing on mathematical tools from network science and graph theory. However, these two different views are rarely addressed simultaneously, largely because few tools exist that account for patterns \emph{between} nodes while simultaneously considering activation \emph{of} nodes. We address this gap by developing a technique that can be used to uncover groups of brain regions (nodes) that are both functionally connected (edges) and share similar activation magnitudes (encoded by annotations on each node). Specifically, we solve the problem of community detection on weighted networks with continuous (non-integer) annotations by deriving a generative probabilistic model. This model generates communities whose members connect more densely to nodes within their own community than to nodes in other communities, and whose members share similar annotation values. We demonstrate the utility of the model in the context of neuroimaging data gathered during a motor learning paradigm, where edges are task-based functional connectivity and annotations to each node are beta weights from a general linear model that encoded a linear decrease in blood-oxygen-level-dependent signal with practice. Interestingly, we observe that individuals who learn at a faster rate exhibit the greatest dissimilarity between functional connectivity and activation magnitudes, suggesting that activation and functional connectivity are distinct dimensions of neurophysiology that track behavioral change. More generally, the tool that we develop offers an explicit, mathematically principled link between functional activation and functional connectivity, and can readily be applied to a other similar problems in which one set of imaging data offers network data, and a second offers a regional attribute.}
\end{abstract}

\end{frontmatter}

\newpage







\section{Introduction}

\begin{figure}
\includegraphics[]{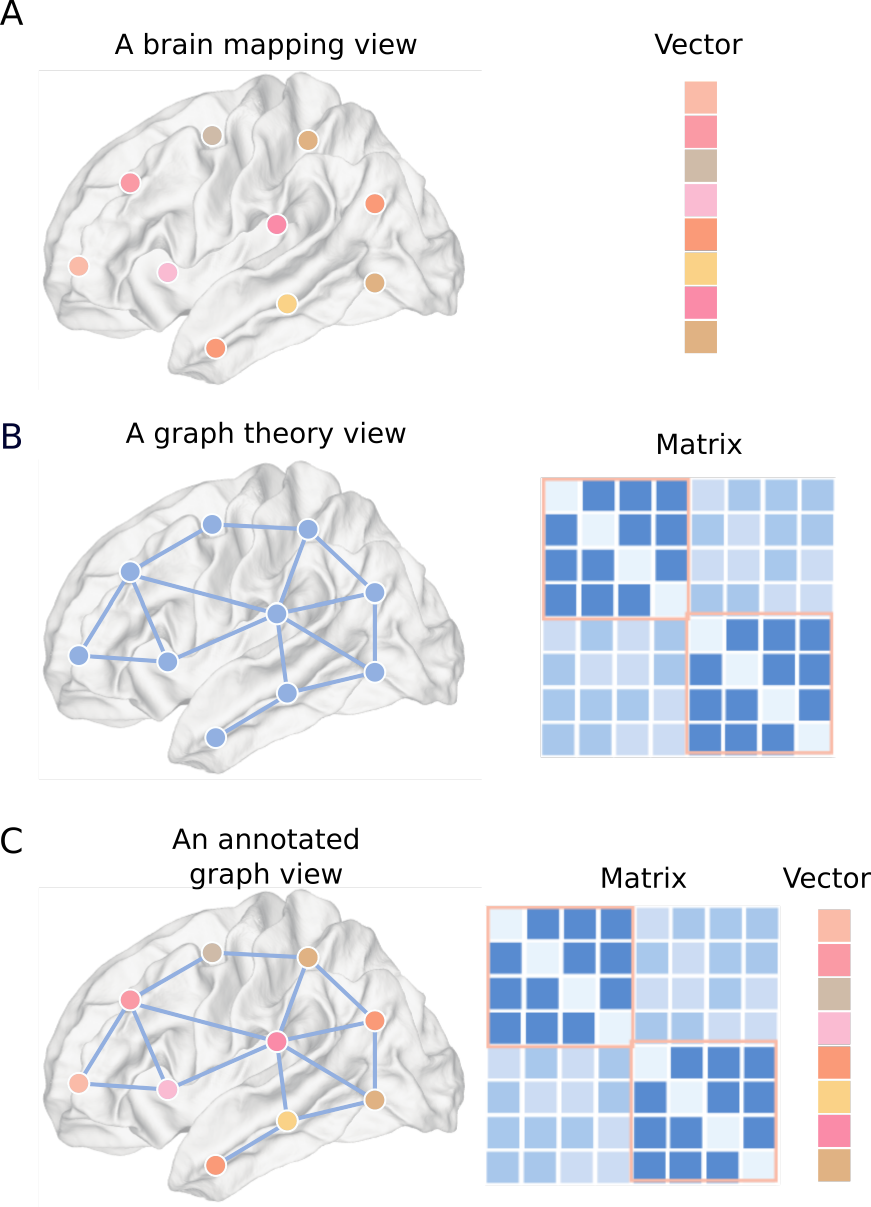}
\caption{\textbf{Philosophical Perspectives on Brain Function.} \emph{(A)} In the field of brain mapping, cognitive function is thought to be driven largely by the magnitude of activation in specific brain areas. From a mathematical perspective, this corresponds to representing brain function as a vector of activation magnitudes across regions of interest. \emph{(B)} In the field of graph theory, cognitive function is thought to be driven by the pattern of connectivity between brain regions. From a mathematical perspective, this corresponds to representing brain function as an adjacency matrix, hard coding the relationships between brain regions. \emph{(C)} Annotated graphs facilitate a perspective that bridges these two disparate fields of view, by allowing an assessment of both patterns of inter-regional relationships and patterns of regional features simultaneously. From a mathematical perspective, this corresponds to representing brain function as a combination of both a vector of annotations, and an adjacency matrix hard coding inter-regional relationships.}\label{fig1}
\end{figure}

Since the ages of phrenology, humans have sought to formulate a map of how brain physiology drives human thought and behavior. Indeed, with the advent of noninvasive neuroimaging techniques, the quest to define these maps has guided much of the intellectual effort in large-scale human neuroscience \cite{Poldrack2006,Culham2001,Cabeza2001,Durston2006}. Over the past two decades, these efforts have delineated brain regions that code for specific cognitive functions by examining changes in the magnitude of activity in a brain area during the performance of a given task \cite{Ford2005,Gruber2002,Gould2003} (Fig.~\ref{fig1}A). Yet, while activity magnitudes are clearly important features of neurophysiological signals, recent evidence suggests that cognitive functions may also be parsed from one another via differences in the patterns of connectivity between many brain areas \cite{Bassett2009,Supekar2013,Moseley2015,Sarpal2014} (Fig.~\ref{fig1}B). Such patterns are parsimoniously defined and studied in the emerging field of \emph{network neuroscience}, which draws on mathematical tools from network science and graph theory.

The question of whether cognition is quintessentially a result of activity patterns or a result of connectivity patterns is perhaps a misleading one. In reality, activity and connectivity are two complementary measures of the same underlying physiology, offering related \cite{Bassett2011b} but not redundant \cite{Bassett2015} information. Despite their theoretical complementarity, the two methods of studying neuroimaging data are rarely used in concert, challenging the construction of holistic theories of neurophysiological mechanisms. This lack of intellectual integration is largely due to the fact that few tools exist that account for patterns \emph{between} nodes while at the same time considering values \emph{on} nodes. In fact, the lack of these tools also hampers progress in other contexts that require the fusion of multimodal neuroimaging data, such as understanding the relationships between grey matter volumes linked by structural connectivity, or how spatial distributions of FDG PET are linked to patterns of functional connectivity.

We address this limitation by developing tools for community detection that are directly applicable to problems of the form in which elements in a network possess attributes (Fig.~\ref{fig1}C). However, to be concrete, we focus our exposition and validation of these tools on linking regional activity and connectivity. The specific method we propose here utilizes graph theory to reformulate the data as an annotated graph \cite{Yang2014,Zhang2015,Bothorel2015,Hirc2014} in which each node in the graph has an associated annotation. This formulation enables us to uncover fundamental organizational principles in the network by identifying network communities. Importantly, communities within an annotated network are a function of both the network and the annotations. We formulate a model in which the members of detected communities will be more densely connected to nodes within their own community than to nodes in other communities, and they will also tend to have similarly valued annotations.

When considering the link between regional brain activity and connectivity, the underlying network of the annotated graph will be a functional connectivity network, and the annotations will incorporate information about the activity level of each brain region. By our construction, the incorporation of the annotations allows more control over the communities detected by biasing the communities with the regional activity magnitudes, making certain partitions of brain regions into functional communities more likely than others. We hypothesize that the degree to which functional connectivity and activity magnitude are related to each other is an important and previously un-probed feature of neurophysiological dynamics. Specifically, information contained in both types need not be fully redundant. Therefore, considering both simultaneously will increase our information about the system. Moreover, we hypothesize that differences in the relationship between activity and connectivity at the meso-scale may be detectible in individual subjects. Finally, we hypothesize that the mapping of activity and connectivity during task performance will be linked to individual differences in learning, such as that evident in motor skill acquisition.

To test these hypotheses, we acquire task-based fMRI data from 20 individuals as they perform a discrete sequence production (DSP) task over the course of 6 weeks of practice \cite{Bassett2015,Wymbs2015}. We summarize these data in functional brain networks where nodes represent 112 brain regions delineated by the Harvard-Oxford atlas, and where edges represent a set of estimates of functional connectivity between brain regions in 1--2 minute windows during task performance. We annotate these networks such that each node is also associated with the sum of a set of weighted beta estimates from a general linear model, one for each learning stage, that represents a decreasing linear slope of regional activation with practice. Here, we have made the choice to select annotations that encode a change in activity over the course of learning, rather than solely activation level. This decision is motivated by recent evidence suggesting that change in activity level is more suited than pure activity level in investigating learning performance and differences in individual learning \cite{Wymbs2015}. 

Using these annotated graphs, we develop a community detection method \cite{Newman2006,Lancichinetti2009,Newman2002,Newman2004b} for weighted graphs with continuous annotations. Using this method, we are able to reconcile two distinct yet parallel streams of information to determine how change in activity and connectivity relate to each other during learning. It has recently been shown that individual differences in connectivity explain individual differences in learning \cite{Bassett2015}. Additionally, it has been suggested that the decrease in activity during learning may be a generic effect that does not sufficiently explain individual differences in rate or quality of learning \cite{Wymbs2015}. Therefore, if this is true, we hypothesize that the two streams of information will become less similar as connectivity profiles evolve with increased learning. We apply our method to the data to identify, from the set of 112 regions, groups of brain regions that are both densely functionally connected with one another and that share similar beta weights with one another. We validate the method with statistical testing and describe the role of model parameters on expected results. In applying this method to real data, we show that dissimilarity between functional connectivity (the network) and decrease in BOLD activation (the annotations) is correlated with higher rates of learning in this visual-motor task. These results support the utility of this approach in linking studies of activity and connectivity, and more generally offer a mathematical framework to assess relationships between regional attributes and inter-regional relationships across other multimodal contexts.

\begin{figure*}
\centerline{\includegraphics{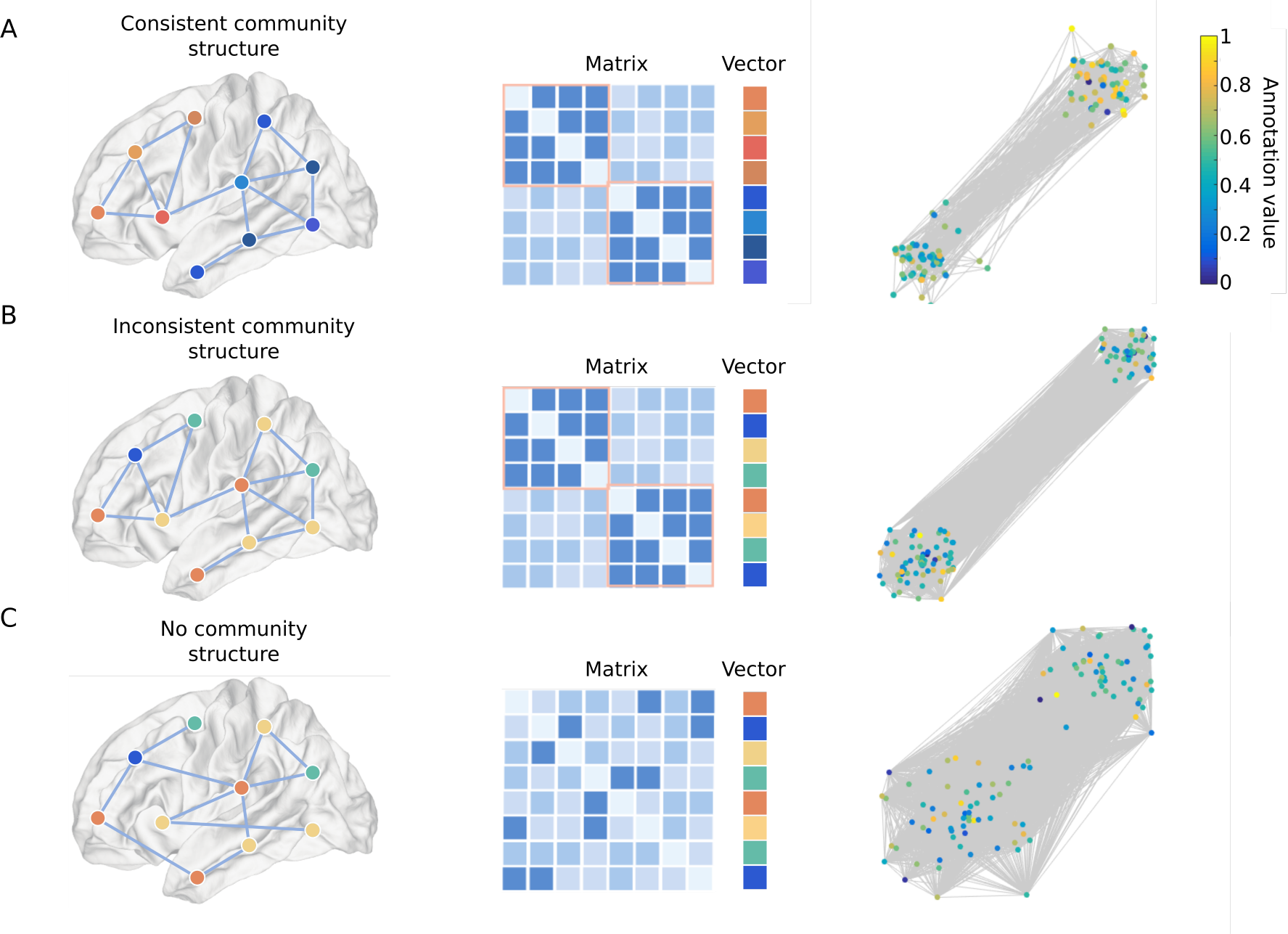}}
\caption{\textbf{A Schematic of the Analytic Procedure: Community Detection on Annotated Brain Graphs.} The left column visually depicts an annotated network where nodes are connected to each other, and colors represent activation magnitudes (annotation value), the middle column shows this network as an adjacency matrix encoding the inter-regional connectivity patterns accompanied by a vector encoding activity magnitudes, and the right column shows the communities resulting from our model. \emph{(A)} An example of a community structure resulting from a functional network and annotation that have consistent structure. Nodes that are densely connected to each other also have similar annotation values. The resulting detected communities reflect this consistency. \emph{(B)} When the annotation structure does not match the network structure, this method will ignore annotations and the resulting community structure will be solely dependent on the network interconnections. The resulting community members will be densely interconnected, but not share similar annotations. \emph{(C)} When no densely interconnected groups of nodes exist, all nodes are assigned to a single community.}\label{fig2}
\end{figure*}

\section{Materials and Methods}

\subsection{Experimental Design}

Twenty-two healthy volunteer subjects (9 male, 13 female, average age 24) provided written informed consent according to the IRB of the University of California, Santa Barbara. None of the volunteers had any history of neurological or psychiatric diseases. One subject was excluded due to excessive head motion, and one subject did not complete the full experiment. Prior studies utilizing this data include \cite{Bassett2015,Wymbs2015}.

The DSP task training was performed over a period of six weeks with four MRI training sessions at the beginning of week 1, and the end of weeks 2, 4 and 6.  On the first day of the experimental procedure, the subjects underwent an initial MRI training session and were given and taught how to use an electronic training program at home. The subjects were required to train for at least 10 of the 14 days between MRI sessions. In-home training sessions consisted of subjects practicing a set of 10 element sequences with their right hand. Sequences were presented via a horizontal arrangement of stimuli, and subjects were instructed to press the corresponding key on a keypad (Fig.~\ref{fig3}). A new stimulus was displayed directly following a correct key press. An incorrect key press resulted in the pausing of the sequence until the correct response was given with no time limit.

Practice trials began with a sequence-identity cue that corresponded to one of six possible sequences in a pseudo-randomized order, each of which was presented at one of three possible training levels (Table~\ref{levels}). Two sequences were presented as rarely trained, two as moderately trained, and two as extensively trained, each sequence was practiced for 10 trials per session. At the end of the 10 trials, subjects received information about their error rate and time to completion.

During the four MRI training sessions, subjects were given identical sequences as they practiced at home, but all sequences were presented with 50 trials. Here we consider 10 different learning stages (stages A-J), where stage A is the average of the first three experimental learning stages, and stages B-J are experimental stages 4-12 (Table~\ref{levels}). Stages are numbered according to training trials. Thus stage 1-3 each have 50 training trials, stage 4 has 110, and stage 12 has 2120.

\subsection{Estimating Learning Rate}

For each sequence, we recorded the \emph{movement time}, which is defined as the time between the first and last button press. For a sequence of a single type, movement time was plotted against trial number and a double exponential was fit to the curve. The fit was constructed to be robust to outliers (fit.m in MATLAB Curve Fitting Toolbox with option 'Robust', type 'LAR'). The fit equation was of the form $TC=D_1e^{-t\kappa}+D_1e^{-t\lambda}$ where $D_1$, $D_2$ are positive constants, and $\kappa$ is related to the steepness of the learning curve such that larger values correspond to faster learners.

\subsection{Imaging Acquisition}

Scans were performed at the University of California, Santa Barbara on a 3T Siemens TIM Trio scanner with a 12-channel phased-array head coil. For whole brain scans, a single-shot echo planar imaging sequence was used that was sensitive to BOLD contrast. Acquired scans were 37 slices per repetition time (2000 ms TR, 3 mm thick, 0.5 mm gap). Scans had an echo time of 30 ms, the flip angle was $90^o$, the field of view was 192 mm, and the acquisition matrix had dimensions 64$\times$64. Structure high-resolution T1-weighted whole brain images (15 ms TR, 4.2 ms echo time, $90^o$ flip angle, 256 mm field of view, 0.89 mm slice thickness, 256$\times$256 acquisition matrix) were also collected for each subject.

Data preprocessing was conducted with the Statistical Parametric Mapping tool (SPM8, Wellcome Trust Center for Neuroimaging and University College London, UK). The structural T1 image was normalized to the MNI-152 template (3$\times$3$\times$3 mm). Subsequently, the raw functional data was realigned and coregistered to the normalized T1 structural image. The functional images were smoothed with a Gaussian filter (isotropic, 8 mm FWHM).

\subsection{Network Construction}

We parcellated the brain into 112 cortical and subcortical regions of interest using the Harvard-Oxford atlas. Time series were extracted for each region of interest over the course of each scan. Functional connectivity between regions was determined by a coherence between wavelet coefficients extracted using a continuous wavelet transform \cite{Bassett2011,Zhang2016}. Here we chose to use wavelet scale 2, corresponding to the frequency band 0.06--0.125 Hz. This method allows for the detection of signal changes in a non-stationary time series, and is particularly useful in the context of noise. Coherence between these wavelet coefficient time series produced an $N \times N$ matrix representing the pairwise functional connections between regions of interest.

\subsection{Stochastic Block Model Construction}

The method used to detect communities here is a variant of the weighted stochastic block model \cite{Aicher2014,Aicher2013}. In general, both weighted and unweighted stochastic block models are extensions of traditional community detection techniques \cite{Porter2009,Fortunato2010}, which allow for the identification of meso-scale structure in networks including but not limited to network communities or modules \cite{Karrer2010}. 

To extract network communities, we employ a stochastic block model \cite{Mossel2012,Abbe2015b,Decelle2011,Abbe2015c}, which is a generative network model of random graphs that tends to produce networks composed of densely interconnected nodes that form communities, and communities that are sparsely connected to each other. We leverage this property to detect network communities by finding the maximum likelihood estimate parameters of the stochastic block model network that most closely resembles the real network. 

To better appreciate the construction of the weighted stochastic block model, we first review the construction of the standard binary stochastic block model, which was first introduced by Holland \cite{Holland1983} for the study of mesoscale structure in binary graphs \cite{Karrer2010}. In this method, a node $n$ is assigned to community $k$, and edges are placed independently from this node to all other nodes in the network depending only on the group membership of the two nodes. Following \cite{Aicher2013,newman2015structure}, the un-weighted stochastic block model can be written as:
\begin{equation}
\textup{Pr}\left ( A|z,\theta  \right )=\prod_{ij}\theta _{z_{i}z_{j}}^{A_{ij}}\left ( 1-\theta_{z_{i}z_{j}}  \right )^{1-A_{ij}},\label{SBM}
\end{equation}
where $z_{i}$ is the community assignment for node $i$, $A$ is the adjacency matrix of the network, $A_{ij}$ is the binary edge weight between nodes $i$ and $j$, and $\theta_{ij}$ is the probability of an edge existing between community $i$ and community $j$.

The traditional stochastic block model described above is able to accommodate networks with only binary edges. Often times, including in the current application, networks have fully weighted edges, and the process of binarizing edges in order to form a binary network is undesirable. Therefore, we formulate a method of community detection that is valid on weighted networks. The stochastic block model can be extended to include weighted edges following \cite{Aicher2014}. This is a particularly beneficial extension in the context of neuroimaging data, where the use of weighted networks \cite{Rubinov2011,Bassett2016small} avoids the need to binarize the graph via thresholding \cite{Bassett2011b,Garrison2015instability,Ginestet2011brain}. Thresholding often involves keeping only the strongest connections, and ignores information contained in weak connections, which has previously been shown to be valuable \cite{Bassett2011b,Santamecchi2014,Cole2012,Schneidman2006}. In the case of the stochastic block model, we can define the probability of a graph being generated from this model as:
\begin{equation}
\textup{Pr}\left ( A|z,\theta  \right )\propto\textup{exp}\left ( \sum_{ij}T\left ( A_{ij} \right ) \cdot \eta\left ( \theta_{z_{i}z_{j}} \right )\right ),\label{WSBM}
\end{equation}
where the probability can be expressed as a distribution belonging to the exponential family. Here $T$ is the sufficient statistic of the distribution, and $\eta$ is the natural parameter.

\subsection{Continuously Annotated Weighted Stochastic Block Model}

The networks studied in the current work are both weighted and annotated. We now have a method to detect communities on the weighted networks, but this method does not yet incorporate any information from the network annotations. Having defined a generalizable model to identify meso-scale structure in weighted networks, we next wish to add continuously valued annotations to the network and use these annotations to assist in deriving community structure. Rather than computing $\textup{Pr}\left ( A|z,\theta  \right )$ of the graph, where the probability is dependent only on the adjacency matrix, we would like to calculate $\textup{Pr}\left ( A,x|z,\theta  \right )$ where we now consider the graph annotated by continuous annotation $x$. A solution to this problem has previously been developed for the case of continuously valued annotation placed atop a binary graph architecture \cite{newman2015structure}. To assume independence between the annotation $x$ and the network graph $A$, we re-write the equation as:
\begin{equation}
\textup{Pr}\left ( A,x|z,\theta  \right )=\textup{Pr}\left ( A|z,\theta  \right )\textup{Pr}\left ( x|z,\theta  \right ),\label{mWSBM}
\end{equation}
where the first term of the right side of the equation is simply the standard weighted stochastic block model derived in \cite{Aicher2013,Aicher2014}, and the second term accounts for continuously valued annotation established in \cite{newman2015structure}.

Here, we will describe how to calculate the annotation term in the above equation. The handling of continuously valued annotation assigned to binary graphs was previously developed by Clauset in \cite{newman2015structure}. We adopt this method and apply it to the weighted stochastic block model to reach the final form of our model. Following \cite{Aicher2014} and \cite{newman2015structure}, we continue the derivation as follows:
\begin{equation}
\textup{Pr}\left ( A,x|z,\theta  \right )= \textup{Pr}\left ( A|z,\theta  \right )\textup{Pr}\left ( x|z,\theta  \right )=\textup{Pr}\left ( A|z,\theta  \right )\textup{Pr}\left ( x|z \right ).
\end{equation}
From \cite{Aicher2013} we have that
\begin{equation}
\textup{Pr}\left ( A|z,\theta  \right )=\prod_{i<j}h(A_{ij})\textup{exp}\left ( T(A_{ij}) \cdot \eta(\theta)\right ),
\end{equation}
and from \cite{newman2015structure} we have that
\begin{equation}
\textup{Pr}\left ( x|z  \right )=\prod_{i}\sum_{j=0}^{P}\gamma_{z_{i}j}B_{j}(x_i),
\end{equation}
where $B_{j}(x)$ are the Bernstein polynomials of degree $P$, and $B_{j}(x)=\binom{P}{j}x^j(1-x)^{P-j}$ for $j = 0...P$. Bernstein polynomials are an appropriate basis set due to the fact that they fall between 0 and 1, and $\sum_{j=0}^{N}B_j\left ( x \right )=1$ \cite{newman2015structure}. Here $\gamma$ is a normalization term $\sum_{z}\gamma_{zj}=1$ so that $\sum_{z}\textup{Pr}(z|x)=1$. This gives the full form of the weighted stochastic block model with continuously valued annotations as
\begin{equation}
\textup{Pr}\left ( A,x|z,\theta  \right )= \\
\prod_{i<j}h(A_{ij})\textup{exp}\left ( T(A_{ij}) \cdot \eta(\theta)\right ) \\
\prod_{i}\sum_{j=0}^{P}\gamma_{z_{i}j}B_{j}(x_i).\label{CWSBM}
\end{equation}
This must be modified slightly to reach the final form, due to the fact that $\textup{Pr}\left ( A|z,\theta \right )$ has $n^2$ terms for a full matrix, but $\textup{Pr}\left ( x|z \right )$ has only $n$ terms, where $n$ is the number of nodes. We therefore perform a correction by exponentiating the second term by $n$ to reach the final form
\begin{equation}
\textup{Pr}\left ( A,x|z,\theta  \right )= \prod_{i<j}h(A_{ij})\textup{exp}\left ( T(A_{ij}) \cdot \eta(\theta)\right )  \left [  \prod_{i}\sum_{j=0}^{P}\gamma_{z_{i}j}B_{j}(x_I)\right ]^{\alpha n}.\label{CWSBM}
\end{equation}
Here we have also added the tuning parameter $\alpha$ to enable explicit control of the weight of the annotation term. Now, the log likelihood of the two terms will be the same order of magnitude (see Eqn. \ref{logLike}).

Above, we have the final form of the generative model to describe a weighted stochastic block model random network with continuously valued annotations. The next task to complete is to calculate the parameters $(z, \theta)$ that generate a stochastic block model network that most closely resembles our observed network. Under binary conditions, $z$ and $\theta$ could be solved by maximizing the likelihood of this equation by taking derivatives with respect to both $z$ and $\theta$, setting the equation equal to 0, and solving the parameters. However, as discussed in \cite{Aicher2014}, because we desire to work with a weighted network, as opposed to a binary network, solutions may be degenerate. Aicher and Clauset \cite{Aicher2014} address this problem by utilizing Bayesian regularization by considering our parameters $\theta$ and $z$ as random variables and assigning a prior distribution $\pi\left ( z,\theta \right )=\textup{Pr}\left ( z,\theta \right )$ to these random variables. The authors then use this construct combined with Bayes' law to define the posterior distribution $\pi^{*}\left ( z,\theta \right )\propto\textup{Pr}\left (A| z,\theta \right )\pi\left ( z,\theta \right )$.

Next, we will describe how to use this posterior distribution in order to solve for the maximum likelihood estimate of the stochastic block model. The posterior distribution may be calculated directly, but we choose to implement the same simplifying assumption as proposed by Aicher and Clauset. We approximate $\pi^{*}\left ( z,\theta \right )$ as a factorable distribution $q\left ( z,\theta \right )=q\left ( z \right )q\left ( \theta \right )$. The term $q$ is then solved for by maximizing the functional lower bound on the equation derived from the Kullback-Leibler (KL) divergence between the approximation $q$ and the KL posterior. This will ensure that our approximation is close to the real solution. The equation to be maximized in the instance of a weighted matrix with continuous annotation is
\begin{equation}
\mathcal{G}=\mathbb{E}_{q}\left ( \mathcal{L} \right )+\mathbb{E}_{q}\left ( \textup{log} \frac{\pi(z,\theta)}{q(z,\theta)} \right ),
\end{equation} where
\begin{equation}
\mathcal{L}=\sum_{i<j}h\left ( A_{ij} \right )\textup{exp}\left ( T\left ( A_{ij} \right )\cdot \eta\left ( \theta \right ) \right )+\alpha n \sum_{i}\textup{log} \sum_{j=0}^{P}\gamma_{z_{i}j}B_{j}\left ( x_i \right ). \label{logLike}
\end{equation}
Here $\mathcal{L}$ is the log-likelihood of the model, and $\mathbb{E}_{q}$ denotes the expectation value. Finally, we can establish the equation to be maximized.
\begin{multline}
\mathcal{G}=\sum_{i<j}\textup{log}~h\left ( A_{ij} \right )+\sum_{r}\left ( \left \langle T \right \rangle_r + \tau_0-\tau_r\right )\left \langle \eta \right \rangle_r
\\
+\alpha n \sum_{i}q_i(z_i)\textup{log} \sum_{j} \gamma_{z_{i}j}B_j(x_i)+\sum_{r}\textup{log}\frac{z(\tau_r)}{z(\tau_0)}
\\
+\sum_{i}\sum_{z_i}\mu_i(z_i)\textup{log}\frac{\mu_0(z_i)}{\mu_i(z_i)}.\label{G}
\end{multline}
Following Clauset, we have also made the substitution
\begin{equation*}
T_r=\sum_{i,j:\mathcal{R}(z_i,z_j)=r}T(A_{i,j}),
\end{equation*}
where $T_r$ is now the sufficient statistic for all edges between communities $z_i$ and $z_j$. Angle brackets indicate expectation values. The variables $\mu$ and $\tau$ are our new distribution parameters, where we impose the prior $\mu_0(k)=1/k$, where $k$ is the number of communities. Physically, $\mu_i(k)$ can be interpreted as the probability of node \emph{i} belonging to community \emph{k}.

Because we impose the fact that our distribution must be exponential, the forms of the sufficient statistic $T$ and natural parameter $\eta$ are known \cite{Aicher2014}:
\begin{equation*}
\left \langle T \right \rangle_r=\sum_{ij}\sum_{\mathcal{R}(z_i,z_j)=r}\mu_i(z_i)\mu_j(z_j)T(A_{ij}),
\end{equation*}
\begin{equation*}
\left \langle \eta \right \rangle_r=\frac{\partial \textup{log}Z(\tau_r)}{\partial \tau_r}.
\end{equation*}

Now, taking derivatives with respect to $\mathcal{G}$ will lead us to the update rules of our solution:

\begin{equation}
\frac{\partial \mathcal{G}}{\partial \tau_r}\propto\left \langle \tau \right \rangle_r+\tau_0-\tau_r\label{updatetau},
\end{equation}

\begin{equation}
\frac{\partial \mathcal{G}}{\partial \mu_i}\propto \textup{exp} \left [  \alpha n~\textup{log}\sum_{j}\gamma_{z_{i}j}B_{j}(x_i)\right ]~\textup{exp}\left ( \sum_{r}\frac{\partial \left \langle T \right \rangle_r}{\partial \mu_{i}(z)}\cdot \left \langle \eta \right \rangle_r \right ).\label{updatemu}
\end{equation}
Lastly, we require the update equations governing $\gamma$, which have been previously derived in \cite{newman2015structure} to be:
\begin{equation}
\gamma_{sj}=\frac{\sum_{u}\mu_u(k)Q_{j}^{su}}{\sum_{tu}\mu_u(t)Q_{j}^{tu}},
\end{equation}
\begin{equation}
Q_{j}^{su}=\frac{\gamma_{sj}B_{j}(x_u)}{\sum_{k}\gamma_{sk}B_{k}(x_u)}.
\end{equation}

The implementation of these update rules is shown in pseudocode in Algorithm~\ref{alg}. More generally, the method ensures that the degree to which the annotations influence the output community structure is dependent on (i) a user-defined tuning parameter, and (ii) the underlying generative model. In the latter case, the annotation will have more influence on the community structure if appropriate parameters $\gamma_{ij}$ can be found that will increase the likelihood of Eq.~\ref{logLike}, and is dependent on the assumptions and construction of the model.

\subsection{Estimation of Annotation Contribution}

In the main text, we define the metadata contribution metric, which quantifies the amount that adding metadata to the network alters the calculated community structure. We estimate the degree of contribution of the annotation to a network's community structure with the following equation:
\begin{equation}
\mathcal{Y}=\textup{MI}\left ( x,z \right )-\textup{MI}\left ( x,z_o \right ) \label{MIcontrib},
\end{equation}
\noindent where MI denotes mutual information, \emph{x} are the annotations, \emph{z} is the community structure generated using the current model, and $z_{o}$ is the community structure generated by the standard weighted stochastic block model (WSBM) with no annotations. The quantity $\textup{MI}\left ( x,z \right )$ will increase as the contribution of annotation to the final community structure increases. Subtracting $\textup{MI}\left ( x,z_o \right )$ corrects for any baseline similarity or dissimilarity that may exist between the two. Because community structure $z$ is a discrete vector and $x$ is, by definition here, a continuously valued vector, we use a modified calculation of mutual information \cite{Ross2014}.

\subsection{Comparing Partitions of a Network Into Communities}

During the validation process, we calculated the similarity of partitions generated with unpermuted data with each other, and also calculated their similarity with partitions from the permuted model. We quantified partition similarity by calculating the $z$-score of the Rand coefficient. Here we will specify how to calculate the $z$-score of the Rand coefficient (zrand) as presented in \cite{Traud2011}. This is simply the $z$-score of the Rand coefficient $w$:
\begin{equation}
z_R=\frac{w-\mu_w}{\sigma_w}.
\end{equation}
More explicitly, this statistic can be calculated in the following way:
\begin{equation}
z_{R}=\frac{1}{\sigma_w}\left ( w-\frac{M_1M_2}{M} \right ),
\end{equation}
\begin{multline*}
\sigma^{2}_{w}=\frac{M}{16}-\frac{(4M_1-2M)^2(4M_2-2M)^2}{256M^2}+\frac{C_1C_2}{16n(n-1)(n-2)} \\
+\frac{\left [ (4M_1-2M)^2-4C_1-4M \right ]\left [ (4M_2-2M)^2-4C_2-4M \right ]}{64n(n-1)(n-2)(n-3)},
\end{multline*}

\begin{equation*}
C_1=n(n^2-3n-2)-8(n+1)M_1+4\sum_{i}n^3_{i\cdot},
\end{equation*}
\begin{equation*}
C_2=n(n^2-3n-2)-8(n+1)M_2+4\sum_{j}n^3_{\cdot j}.
\end{equation*}
Here, $n_{ij}$ is the number of nodes assigned to group $i$ in partition 1 and assigned to group $j$ in partition 2, $n_{i\cdot }=\sum_j n_{ij}$, $n_{\cdot j}=\sum_i n_{ij}$, $M_1=\sum_{i}\binom{n_{i\cdot }}{2}$, $M_2=\sum_{j}\binom{n_{\cdot j}}{2}$, $M$ is the total possible number of pairs, and $n$ is the total number of nodes, and $w=\sum_{ij}\binom{n_{ij}}{2}$. A working implementation of this statistic is freely available \cite{NCT}.

\subsection{Implementation}

The implementation of this model was adapted from freely available code from A. Clauset (http://tuvalu.santafe.edu/\textasciitilde aaronc/wsbm/). An outline of the algorithm is provided in Algorithm~\ref{alg}. In this work, the exponential family distribution employed to model edge weights was a Poisson distribution, in line with previous work \cite{Aicher2013}. The model requires the investigator to choose the number of communities $k$ that nodes should be distributed among. In many neuroimaging studies, this number is not known, and therefore we estimated partitions for multiple values of $k$. For further analysis, we choose the $k$ associated with the model that had the highest likelihood. Importantly, the model is non-deterministic and dependent on initialization values; different runs of the algorithm on the same data can produce different community structure estimates. To address this issue, for each network the model was run 50 times. For further analysis, we selected the output of the run with the highest likelihood.

\begin{table}
\caption{\textbf{Training Level by Sequence and Session.} This table depicts the training level number of trials performed for each level of training and session number (adapted from \cite{Bassett2015}).}
\begin{tabular}{@{\vrule height 10.5pt depth4pt  width0pt}lcccc}
\cline{1-5}
  &Naive&Early&Middle&Late\\
\hline
Rare&50&110&170&230\\
Moderate&50&200&350&500\\
Extensive&50&740&1,430&2,120\\
\hline
\end{tabular}
\label{levels}
\end{table}

\begin{table}
\caption{Brain areas in motor and visual modules.}
\begin{tabular}{@{\vrule height 10.5pt depth4pt  width0pt}ll}
\cline{1-2}
  Motor&Visual\\
\hline
L,R precentral gyrus&L,R intracalcarine cortex\\
L,R postcentral gyrus&L,R cuneus cortex\\
L,R superior parietal lobule&L,R lingual gyrus\\
L,R supramarginal gyrus, anterior&L,R supracalcarine cortex\\
L,R supplementary motor area&L,R occipital pole\\
L parietal operculum cortex&\\
R supramarginal gyrus, posterior\\
\hline
\end{tabular}
\label{areas}
\end{table}

\begin{algorithm}
 \KwData{Weighted network $A$, continuous annotation annotations $x$}
 \KwResult{Community assignments of each node to $k$ communities}
 initialization\;
 \While{$\mu$,$\tau$ not converged}{
  \For{$K=1...k$}{
  update $\tau$: $\tau_{r}=\left \langle \tau \right \rangle_{r}+\tau_{o}$\\
  update $\eta$: $\left \langle \eta \right \rangle_r=\frac{\partial }{\partial r}\textup{log}~Z(\tau)$
   }
    \While{$\mu$ not converged}{
    \For{$i=1...n$}{
    update $\mu$: $\mu_i(z)=\textup{exp}\left ( \sum_k \frac{\partial \left \langle T \right \rangle_r}{\partial \mu_i(z)} \cdot \left \langle \eta \right \rangle_r \right )\textup{exp}\left [ \alpha n~\textup{log}\sum_{j}\gamma_{z_{i}j}B_{j}(x)\right ]$\\
    \While{$\gamma$ not converged}{
    update Q: $Q_{j}^{su}=\frac{\gamma_{sj}B_{j}(x_u)}{\sum_{k}\gamma_{sk}B_{k}(x_u)}$\\
    update $\gamma$: $\gamma_{sj}=\frac{\sum_{u}\mu_u(k)Q_{j}^{su}}{\sum_{tu}\mu_u(t)Q_{j}^{tu}}$
    }
    }

 }
 }
 \caption{Outline for determining community structure in annotated networks, adapted from \cite{Aicher2013}. The input of this model is the weighted network $A$, and the annotations $x$. The outputs are the terms $\mu_i(z)$, the probability that a node with annotation value $z$ belongs to community $i$. }
 \label{alg}
\end{algorithm}

\section{Results}

\subsection{Developing a Method to Bridge Regional Characteristics and Inter-Regional Relationships}

To construct a method to bridge regional characteristics and inter-regional relationships, we draw on the mathematics of annotated graphs. Annotated graphs are graphs in which nodes in the network do not only have a set of edges affiliated with them, but also characteristics or features that are not defined in relation to other nodes. Intuitively, one could imagine that nodes might share similar connectivity properties and also show similar features. For example, in resting state fMRI, regional time series characteristics (such as entropy) are correlated with inter-regional connectivity characteristics (such as weighted degree) \cite{Bassett2011b,Zalesky2012}. Similar relationships have been uncovered in MEG data acquired during the performance of memory tasks \cite{Siebenhuhner2013}. In these cases, the edge patterns and annotations provide complementary -- and indeed sometimes even redundant -- information. However, one can also imagine that nodes may be densely interconnected with one another while displaying inherently different internal features. A particularly salient example of this distinction can be found in patterns of functional connectivity and task-based activation magnitudes \cite{Bassett2015}. Here, the edge patterns and annotations can provide non-redundant, and indeed at times orthogonal information about the computations that the brain supports.

An important question in any new neuroimaging data set is whether (and to what degree) community structure is present in the network edges, and whether this community structure relates to other features of interest present in the network nodes. To address this question, we develop and employ an annotated expansion of a weighted stochastic block model \cite{Aicher2013,Aicher2014,Jog2015,Zhao2015}, which enables us to explicitly identify communities driven both by dense interconnectivity and by similarities in regional characteristics. Moreover, we employ appropriate null model comparisons, where for example the community structure is destroyed, or the mapping from community structure to regional features is destroyed. For a full description of the mathematics underpinning the model, see Fig.~\ref{fig2} and the Methods section.

\begin{figure*}
\centerline{\includegraphics{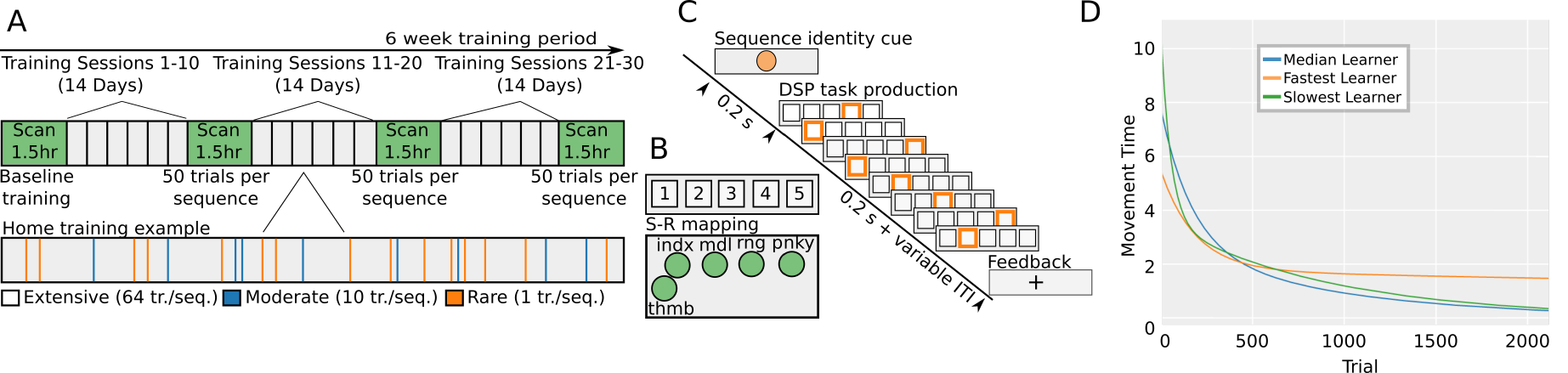}}
\caption{\textbf{A Schematic of the Experimental Design for Motor-Visual Skill Acquisition in the Form of a Discrete Sequence Production Task.} \emph{(A)} The training paradigm consisted of four scan sessions approximately evenly spaced across a six week period. Each consecutive pair of scans was separated by approximately 14 days, in 10 of which subjects practiced the same finger sequences at home on their laptop computers. \emph{(B)} During each trial, a horizontal array of 5 illuminating squares would indicate which button the subject would press: the left-most button corresponding to the thumb and the right-most button corresponding to the pinky on the subject's right hand. \emph{(C}) During the experiment, subjects viewed a screen on which a cue was presented for each sequence. Each sequence was composed of a series of lights presented to the subject, for which the subject was instructed to press the corresponding button. Subjects were instructed to complete the sequence as swiftly and accurately as possible. At the end of each sequence, a fixation cross was displayed for a brief inter-trial interval (ITI). \emph{(D)} To quantify learning rate, we fit a double exponential function to the movement time, defined as the time between the first button press and the last button press of a given sequence, as a function of trial number (see Methods). Fits are shown for the median learner, the fastest learner, and the slowest learner in the cohort.}\label{fig3}
\end{figure*}

\subsection{Behavioral Results}

While the subject was being scanned, data on movement time was also recorded. The rate of learning for each individual was quantified in two different ways: first, the learning rate for each individual over the course of the entire six week trial was estimated (see methods, Fig.~\ref{fig3}D). The relevant feature of this learning rate is the exponential drop-off, rather than the intercept. Second, learning over the course of each trial was recorded. The rate of learning remained relatively constant over each trial.

\subsection{Identifying Communities in Weighted Graphs with Continuously-Valued Annotations}

An explicit goal of our method is to identify groups of brain regions that are densely functionally interconnected with each other, and that also share similar annotations. Consistent with the graph theory literature, we refer to the continuously valued annotations that we use as ``metadata''.  In an initial evaluation of the performance of our method, we seek to quantify the degree to which the communities that we identify are composed of brain regions with similar values of the annotation, which in this case is a weighted set of beta estimates from a general linear model (GLM) that represents a linear decrease in blood-oxygen-level-dependent (BOLD) signal with increasing task practice on a novel motor skill. In essence, we seek to quantify the similarity between a partition of nodes into communities (a vector of categorical values) and the beta weights of the GLM (a vector of continuous values). To calculate this similarity, we use a measure of mutual information developed explicitly for the purpose of measuring similarity between discrete and continuous data sets \cite{Ross2014} (see Methods).

We observe that across sequence types, scanning sessions, and participants, the average mutual information between the annotation and the regional allegiance to communities is $MI=0.1098$ ($SD = 0.1678$).  To interpret the statistical significance of these results, we compare the observe mutual information values with those expected in a non-parametric permutation-based null model. Specifically, we permute the regional labels of the annotation values uniformly at random across all brain regions. Then, we re-apply our method to identify a null set of community assignments. Importantly, these null model partitions were very different from those obtained in the real data. The partition similarity of pairs of real partitions, calculated via the $z$-score of the Rand coefficient \cite{Traud2011}, was significantly larger than the partition similarity of pairs composed of one real partition and one null model partition: paired $t$-test $t=-249.25$, $p=5.27\times 10^{-96}$. We then calculate the mutual information between this null model community structure and the original unpermuted annotations. We observe that the mutual information between the true annotation and the regional allegiance to communities is significantly greater than that expected in the null model (paired $t$-test $t=27.91$, $p=1.46 \times 10^{-148}$). These results confirm that the communities identified by our method are directly informed by the annotation input, being composed of regions that are not only densely functionally interconnected but that also share similar changes in BOLD magnitude with training.

\subsection{Effect of Model Parameters}

The results reported in the previous section are based on a single set of model parameters. Here we describe the role of each of these parameters in the model, discuss a few important considerations in choosing these parameters, and highlight reasonable choices in the context of the data we study here.

\begin{figure}
\centerline{\includegraphics{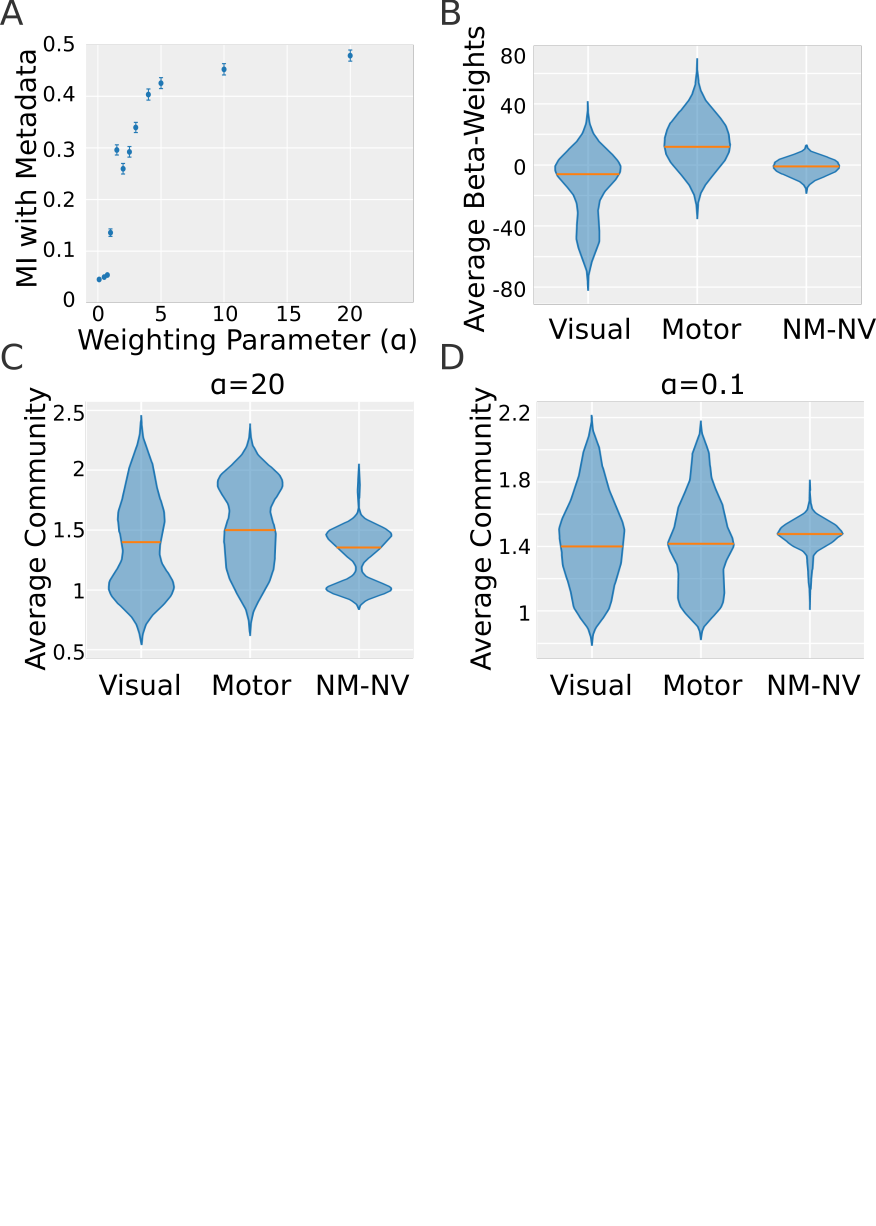}}
\caption{\textbf{Tuning the Sensitivity of the Model to the Annotation.} \emph{(A)} Mutual information between the input annotation and the output community structure increases as the relative weight of annotation in the model is increased. Error bars represent $95\%$ CI. \emph{(B)} The mean annotation values across all 20 subjects by system according to Table~\ref{areas}. \emph{(C)} The average community assignments by system for $\alpha = 20$. As $\alpha$ is tuned up, the community structure approximates the annotation structure. \emph{(D)} The average community assignments for $\alpha = 0.1$. For low $\alpha$, very little of the annotation structure is reflected in the community assignments.}
\label{fig5}
\end{figure}

First, we note that within the model construction, we include a parameter that tunes the importance of the annotation values relative to the importance of the connectivity values in determining the community structure. This parameter is $\alpha$ in Eq.~\ref{CWSBM}, which we will refer to as the weighting parameter.  Intuitively, if the range of annotation values is bounded in [0,1], and the range of the connectivity values is bounded in [0,1], then one potential reasonable choice for the weighting parameter is unity, hard-coding the assumption that the annotations and connectivity provide equal weight to the solution. However, it is also of interest to understand the dependence of the community structure on this choice. To address this question, we study the community structure as a function of the weighting parameter $\alpha$, which we vary from $0.01$ to $20$. We note that if we increase the relative importance of the annotation values in the model by tuning up $\alpha$, we observe an increase in the mutual information between the annotation values and the identified partition (Fig.~\ref{fig5}A). That is to say, as we increase the weight of the annotation term, communities are more likely to be composed of regions with similar values for the annotation.

Here we note that in general, community assignments are unordered, categorical values. Specifically, for some communities 1, 2, and 3, communities 1 and 2 are equally as different as communities 1 and 3. Therefore, averaging the community assignment is treating a categorical value as an ordered value, which is not valid. However, we note that here we are considering only 2 communities, which represents a special case in which averaging the community assignment becomes a valid quantification of community allegiance. Here we also enforce a constraint to guarantee the same community labels across runs.

To gain a better intuition for the role of the weighting term in the observed community structure, we then studied the relationship between the community structure and the annotation for different values of $\alpha$. The annotations themselves irrespective of the functional connectivity show most positive values (greatest beta weights from the GLM) in the visual system, most negative values in the motor system, and values near zero in the rest of the brain (non-motor, non-visual) \cite{Bassett2015} (Fig.~\ref{fig5}B). Here we divide the cortex of the brain into three non-overlapping groups, a group that is functionally related to motor movements, a group that is functionally involved with vision, and a group for all other parts, consistent with previous works \cite{Bassett2015}. Incorporating the information from functional connectivity, we employed our annotated graph method and selected $\alpha$ corresponding to the highest observed value of mutual information between the annotations and the partition (from Fig.~\ref{fig5}A). The resulting community assignment by system shows a similar map to the annotations themselves (Fig.~\ref{fig5}C). Systems that have different mean annotation values are on average assigned to different communities (one-way ANOVA with a random effect for subject and trial, $F = 7.86$, $p = 0.0049$. Identical treatment with a low tuning parameter (Fig.~\ref{fig5}D) also showed a significant difference in community assignment ($F = 18.19$, $p = 1.3\times10^-8$), but the mean community assignment is much more similar across brain systems. To confirm this observation, the mean community assignment for each region was computed for each subject to form a matrix of dimension $20\times3$ at both $\alpha = 0.01$ and $\alpha = 20$. The correlation between these matrices was not significant (Pearson's $r = -0.1039$ and $p=0.4295$) indicating that the distribution of average community assignments over systems in Fig.~\ref{fig5}C and Fig.~\ref{fig5}D are statistically different. Furthermore, the variance of mean community assignment over systems is significantly higher in the case of $\alpha = 20$ \emph{versus} $\alpha = 0.01$ (0.0627 \emph{versus} 0.0015, paired $t$-test $p=1.6047\times10^{-4})$, indicating that the community assignments of all three systems are more different when the tuning parameter is higher.

\begin{figure}
\centerline{\includegraphics{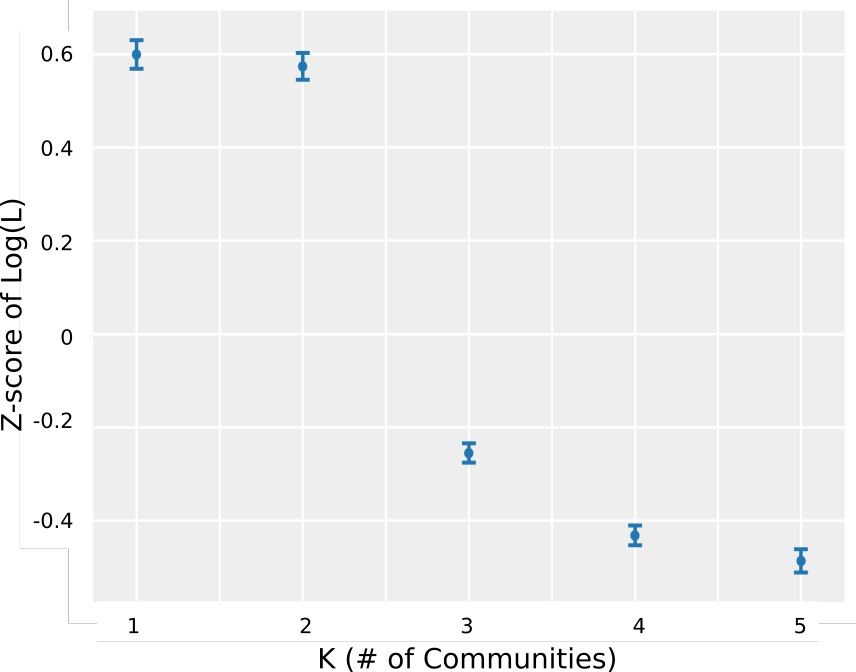}}
\caption{\textbf{Selection of number of communities $k$}. \emph{(A)} This graph shows the mean log likelihood of the generative model as a function of the number of communities $k$. This value has been z-scored across subjects to facilitate better comparison, and demonstrates that $k=2$ is the most likely solution. Error bars represent $95\%$ CI. \emph{(B)} To validate this method of choosing the number of communities $k$, we generated a set of 5 synthetic graphs made up of four communities, and we show one those  graphs here. The color bar indicates the strength of the node-node connection. We then performed community detection on these graphs. \emph{(C)} Our community detection algorithm determines that $k=4$ is the most likely number of communities in these synthetic graphs, as expected.}\label{fig6}
\end{figure}

In addition to the weighting parameter, we must also specify the number of communities $k$ that nodes will be assigned to. While in some scenarios, one might have an \emph{a priori} hypothesis regarding the number of communities present in the data, in many other scenarios, this is not the case. Devoid of \emph{a priori} information, we can proceed in a data-driven fashion \cite{Aicher2013}, using for example a maximum likelihood approach to estimate the likelihood of a solution within the probabilistic model. Specifically, we generate models at different values of $k$ and we select the model with the highest likelihood (Fig.~\ref{fig6} A). In other words, we pick the $k$ associated with the model that has the highest probability of accurately representing the annotated graph. In the context of the data we study here, $k=2$ provides the highest $z$-score of the log-likelihood of the model, averaged over subjects and scans (Note that $k=1$ is a trivial solution and is therefore not examined here). Therefore, all results presented in this text are calculated using $k=2$; making this choice consistent across individuals and scans ensures that individual differences in annotated graph structure are not driven by parameter choice. Then, the job of our model is to decide how to sort the nodes into these two groups, using only activity and connectivity information. We validate this method using graphs that we generated to have 4 communities (Fig.~\ref{fig6} B), and found that this community detection method converged on the correct solution (Fig.~\ref{fig6} C).

\subsection{Relative Independence of Activity and Connectivity During Learning}

Now that we have defined the model, and explored the choice of model parameters, we turn to an evaluation of what insights this tool can provide about the coupling between activity and connectivity in the context of motor skill acquisition. Prior work has demonstrated a complex pattern of increases and decreases in activation profiles during the acquisition of new motor skills \cite{Dayan2011,Wymbs2015,Shannon2016}, and further that these changes can occur over different time scales of learning from slow and deliberate, to fast and automatic. More recent efforts have begun to delineate the patterns of functional connectivity (or graphs) supporting motor learning \cite{Bassett2011,Bassett2013,Bassett2014,Bassett2015,Mantzaris2013,Heitger2012,Xu2014}, and how those patterns are altered following brain injury \cite{Wang2010}.  However, little work has addressed the question of how the observed changes in activity relate to the observed changes in connectivity  \cite{Tomasi2014}. Moreover, evidence in other task and rest contexts has offered conflicting evidence, some in support of a simple and strong relationship between activity and connectivity \cite{Bassett2011b,Zalesky2012,Yu2013} and others in support of a more complex relationship between activity and connectivity \cite{Siebenhuhner2013,Bassett2015}. While these studies address activation magnitude, and here we investigate \emph{changes} in activation magnitudes, these examples nevertheless highlight the need for more investigation into how activation interacts with connectivity.

The annotations we study encode the degree to which activation of brain regions decreased over the course of the 6 week experiment. Our goal is to understand the degree to which these annotations are reflected in patterns of functional connectivity as the skill is acquired. By construction, the algorithm incorporates annotations into the final community structure only if the annotations increase the likelihood of the model \cite{newman2015structure}; annotations will be used to a greater degree if they share structure with the network. This feature of the model allows us to assess the degree of similarity between the functional connectivity network and the regional activity magnitude annotations. If we observe a strong relationship between activity and connectivity, the data would support the notion that the brain areas that decrease in activity form functionally connected communities. On the other hand, if we observe a weak relationship between activity and connectivity, the data would support the notion that the brain areas that increase in activity and the regions that form functionally cohesive modules are complementary but not redundant with one another.

First, we explicitly test for the similarity between the annotations and the functional connectivity. If incorporating the annotations informs the community detection process, then this similarity will be high, while if the annotations do not inform the community detection process, then this similarity will be low.  To estimate this contribution, we test the Pearson correlation coefficient between the degree of each scan-specific network node, and the annotation associated with that node, across all scans acquired during the entire six week trial. Using a Bonferroni-Holm correction for multiple comparisons, none of the $p$-values associated with these correlations were significant ($p>0.05$). These results indicate that the annotation (the beta weights of the GLM) and the network architecture (measured by functional connectivity) were not significantly correlated with each other. Importantly, these data support the notion that brain activity and brain connectivity are complementary but not redundant phenotypes of brain function.

\subsection{Learning Modulates the Activity-Connectivity Relationship}

\begin{figure*}
\centerline{\includegraphics[]{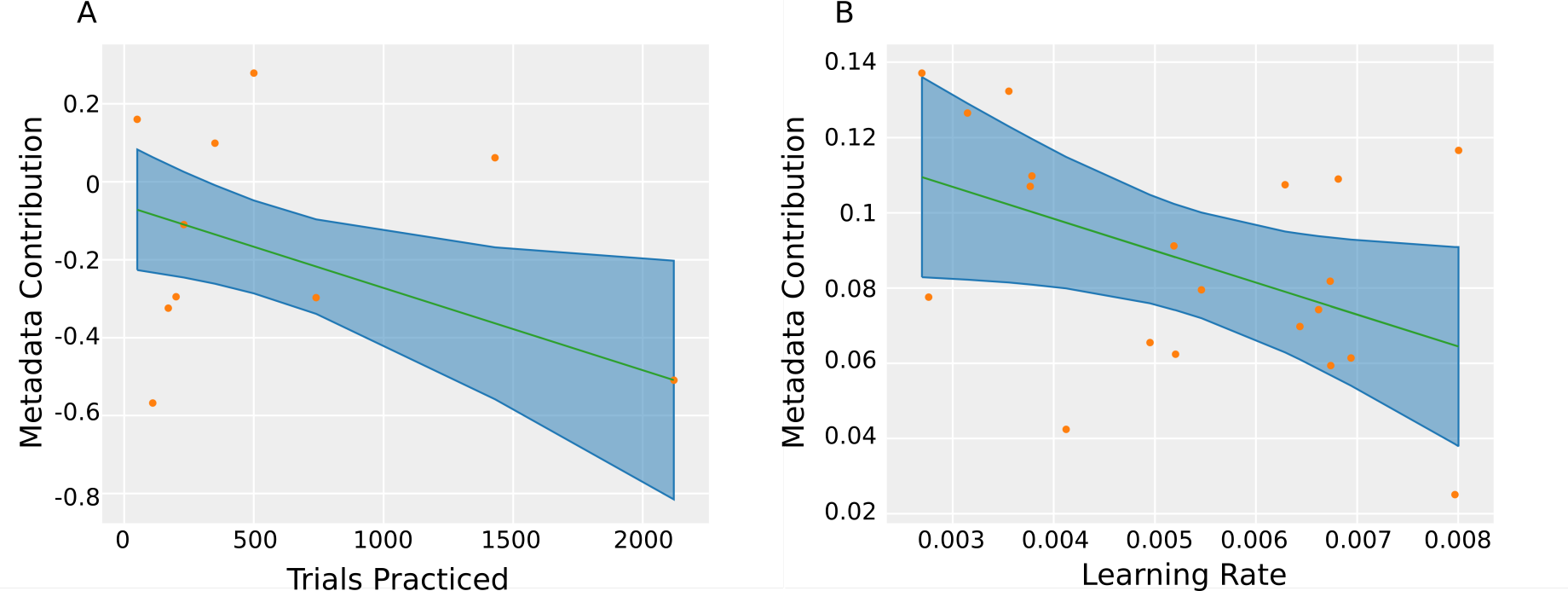}}
\caption{\textbf{Learning as a function of annotation contribution.} \emph{(A)} Annotation contribution as a function of trials practiced ($R^2 = 0.496$, $p = 0.0278$). \emph{(B)} Annotation contribution as a function of learning performance over all 6 weeks ($R^2 = 0.197$ and $p = 0.049$). Each data point represents a subject in the experiment. The line indicates the best linear fit; the shaded area indicates the $95\%$ CI.  }
\label{fig7}
\end{figure*}

Finally, we examine how network structure and the contribution of annotation to community structure (see \emph{Methods}) changes as a function of learning. We calculate the annotation contribution for each scan performed in the experiment. For each of 20 subjects, each scan was binned into one of 10 learning stages (see \emph{Methods} and Table~\ref{levels}), and in this way a $20\times10$ matrix was constructed tabulating the annotation contribution at each learning stage for each subject. To facilitate comparison between subjects, the $z$-score of the $1\times10$ vector for each subject was computed. Fitting a linear model to the median of metadata contribution over subjects for each learning stage \emph{versus} number of trials practiced (from Table~\ref{levels}) revealed a significant negative correlation compared to a constant model ($R^2 = 0.496$, $p = 0.0278$). No significant fit was found when applying this analysis to the null model ($R^2 = 0.007$, $p = 0.877$). This indicates that the contribution of the annotation to the community structure decreases with learning. The fits were calculated using MATLAB's fit.m with the `robust' option to minimize the absolute residual of the fit, ensuring the fit is robust to outliers. These results indicate that activity and connectivity become more autonomous from one another with learning (Fig.~\ref{fig7}A).

Next, we ask whether individual differences in the rate of learning map on to individual differences in the relationship between activity and connectivity. We again estimate the relationship between activity and connectivity using the annotation contribution. To measure learning rate $\kappa$, we calculate the exponential drop-off parameter of the curve of movement time \emph{versus} number of trials practiced (see Methods). Across subjects, we observe a significant negative correlation between annotation contribution (averaged over learning stages) and learning rate $\kappa$ estimated from the full 6 weeks of practice: again, a robust linear fit (MATLAB fit.m option 'robust' set 'on') revealed a significant relationship with $R^2 = 0.197$ and $p = 0.049$ relative to a constant model (Fig.~\ref{fig7}B). Critically, a significant correlation was not observed when using the annotation contribution derived from the non-parametric permutation-based null model ($R^2 = 0.0343$, $p = 0.436$). Furthermore, similar results were observed in the inverse case, when learning rate is taken as the response variable, and metadata contribution is the predictor ($R^2 = 0.245$ and $p = 0.0265$, metadata contribution effect$ = -0.029$), while no relationship was observed in the permuted model ($R^2 = 0.031$ and $p=0.464$). These results indicate that individuals with greater divergence between activity and connectivity profiles are better able to learn than individuals with convergence of activity and connectivity profiles. Importantly, there is no correlation between the mean learning rate (over all subjects) at each stage and the number of trials practiced at that stage ($R^2 = 0.422$ and $p=0.453$).

Finally, we asked whether this inverse relationship is only observed at the coarse level of the entire experiment, or whether it is a feature that is robustly observed over smaller increments of time and training. To address this question, we calculated the learning rate $\kappa$ of each of the 20 subjects at each of the 10 stages of learning (see Methods). We used a linear mixed effects model, allowing intercepts for both subject and number of trials practiced to be represented as random effects, and allowing for the interaction between metadata contribution and subject, as well as metadata contribution and trial. Again, we chose to use a robust fitting method to minimize the absolute residuals. To facilitate equal comparison among subjects, we again chose to $z$-score within each subject across learning stages. Across learning stages, we observed that the annotation contribution was again negatively correlated with individual differences in learning rate (model $R^2=0.44$, metadata contribution effect $=-0.123$, \emph{p} $= 0.007$). In the null model, we observed no significant correlation (model $R^2=0.47$, metadata contribution effect $=0.007$, \emph{p} $= 0.877$). Intuitively, these results indicate that faster rates of learning are associated with systems in which functional connectivity and activity are less aligned.

\section{Discussion}

Here we address the challenge of simultaneously considering functional connectivity and regional activity, an enterprise that is essential to achieve a comprehensive understanding of brain dynamics. To address this challenge, we leverage annotated graphs \cite{Yang2014,Zhang2015,newman2015structure} to formulate a generative probabilistic model that is able to unify these two streams of information. We explore the dependence of the model output on the choice of parameter values, including the weighting parameter $\alpha$ and the number of communities $k$, and we offer a set of best practices for choosing these values in arbitrary data sets. Moreover, we apply the method to functional magnetic resonance imaging data acquired over 6 weeks as subjects learn a new motor skill in the form of a set of finger sequences. We observe that learning is accompanied by a growing segregation of activity and connectivity, and that the individuals who learned best were those with the greatest segregation between activity magnitudes and patterns of functional connectivity. Importantly, these data offer evidence in support of the notion that brain activity and connectivity are complementary but not redundant phenotypes of brain dynamics. More generally, the method that we develop and apply to neuroimaging data is valid for other problems of the form involving a network with nodal attributes.

\subsection{A Method for Linking Node Attributes and Inter-nodal Relationships}

An essential component of our model is that it considers both inter-nodal relationships and nodal attributes. For example, we applied the model to a scenario in which brain regions (nodes) were linked by functional connections (inter-nodal relationships) and also displayed differing beta weights drawn from a GLM (nodal attribute). We use the model to uncover community structure in such annotated graphs, where the partition of nodes into communities is influenced by the annotations \cite{newman2015structure,Hric2016}. We quantify the degree to which annotations are being incorporated into the model using a mutual information constructed explicitly to compare categories with continuous variables. Importantly, the model is tunable in the sense that the weighting parameter $\alpha$ can be used to adjust the relative contribution of the annotations to the identified partition. As this tuning parameter is turned up, the amount of information from the annotation reflected in the community output also increases up until the point at which the community structure approximates the annotations. This allows for explicit control of the composition of the hybrid community structure being generated. The value of $\alpha$ may be selected from prior information, or one could do a search for the $\alpha$ that maximizes the likelihood of the model. To determine the correct number of communities, we build on the method of maximum likelihood \cite{Karrer2010,Abbe2015} due to its simplicity and ease of interpretation. This method offers consistent results, selecting the same number of optimal communities $K$ when tested on over 1000 networks generated under identical experimental conditions \cite{Bassett2015}. Together, the construction of the model and methods for assessment and tuning provide a flexible and generally applicable toolset amenable to problems in which network architectures are accompanied by sets of nodal features.

\subsection{A Model-Based Rather than Correlation-Based Approach}

Relative to a correlation-based approach, model-based approachs offer significant benefits in terms of mechanistic understanding and the possibility of offering a richer explanation for neurophysiological phenomenon. Additionally, the specific model-based approach we develop here offers significant advantages over a correlative approach. For example, we could consider the patterns of functional connectivity and functional activity separately, and identify the community structure in the functional connectivity matrices using a community detection technique or a standard weighted stochastic block model \cite{Holland1983}. Then, we could have calculated the mutual information between the partition of nodes into communities and the annotation. However, because we have not utilized the annotation in the assessment of community structure, the partition that is converged upon can show very little similarity with the annotation. In contrast, a model-based approach explicitly informs the community detection with the annotation values, thereby ensuring that we maintain the greatest possible sensitivity to information shared in both types of data. Moreover, it provides a tuning parameter by which we can systematically assess the similarities and differences between the information housed in the annotation and that housed in the patterns of functional connectivity, a continuous assessment that is simply not possible in the simple correlative approach.

\subsection{Activity and Connectivity as Separate Dimensions of Brain Function}

\begin{figure}
\centerline{\includegraphics[]{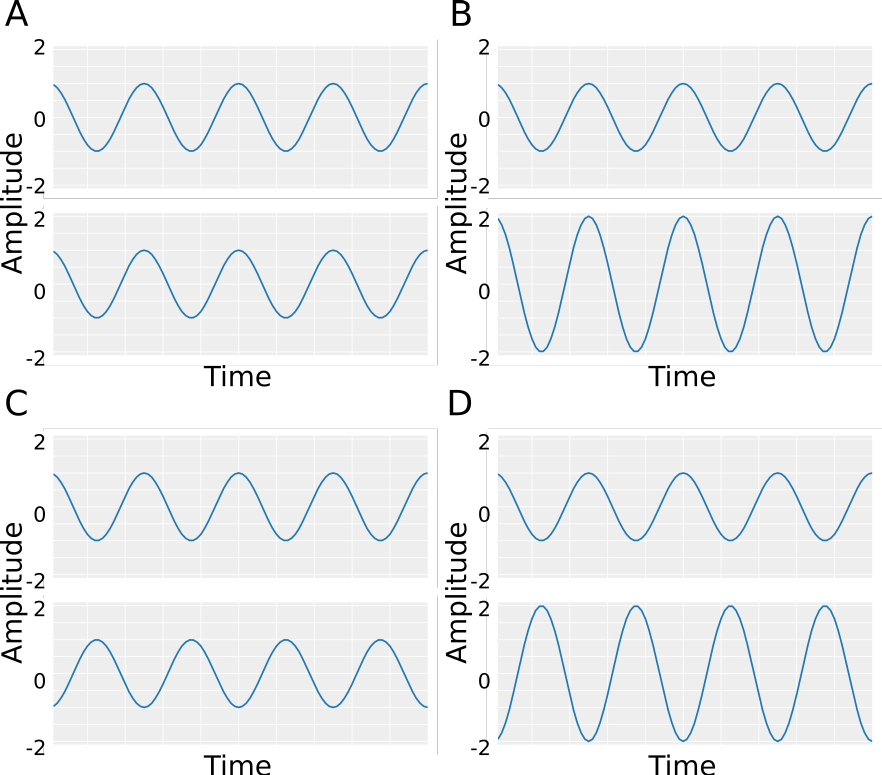}}
\caption{\textbf{Relationship of magnitude and correlation.} Here we demonstrate how the magnitude of a signal and the correlation between time series do not need to be related. \emph{(A)} Two time series of equal magnitude are perfectly correlated ($r=1$), activity and functional connectivity are aligned. \emph{(B)} Two time series of unequal magnitudes are perfectly correlated ($r=1$), activity and functional connectivity are not aligned. \emph{(C)} Two time series of equal magnitudes are perfectly anti-correlated ($r=-1$), activity and functional connectivity are not aligned. \emph{(D)} Two time series of unequal magnitudes are perfectly anti-correlated ($r=-1$), activity and functional connectivity are not aligned.  }
\label{fig8}
\end{figure}

Using this model-based approach, we assess the relationship between task-evoked activation magnitudes and patterns of task-related functional connectivity. The relationship between these two traditionally separate phenotypes of brain dynamics is important when viewed in the historical context of the field of cognitive neuroscience \cite{Gazzaniga2014}.  Traditionally, patterns of task-evoked activation -- as measured by an increase in BOLD magnitudes in task as opposed to rest, or in one type of task \emph{versus} another -- are thought to compose the neural real estate critical for task execution \cite{boynton1996linear}, based on the theory of mental chronectomy \cite{donders1969speed}. Yet, a fundamental understanding of how these two measures might relate to one another (or might be distinct from one another) has remained elusive \cite{Bassett2011b,Zalesky2012}. From a signal processing perspective, it is clear that the magnitudes of time series, and the correlation or coherence between time series need not be at all related to one another (Fig.~\ref{fig8}). This intuition from signal processing, however, can be at odds with the historical intuition from neuroscience, that activation patterns form the fundamental and gold-standard description of brain function, and therefore that when functional connectivity patterns and activation patterns do not align, there must be something wrong. The model-based approach that we develop here offers a principled method by which to study both task-evoked activations and task-evoked functional connectivity. In applying this method to fMRI data acquired during the learning of a new visuo-motor skill, we observe a clear distinction between the information housed in patterns of activity and connectivity: regions that show high beta weights in a GLM testing a linear decrease in activity with task practice do not tend to co-localize in functional network communities. These results complement prior observations of distinctions between characteristics of activity and connectivity \cite{Siebenhuhner2013}, and further motivate additional studies of the relationships between these two measures of brain dynamics in other task scenarios.

\subsection{Relationships Between Activity, Connectivity, and Learning}

Motor learning induces clear changes in both functional connectivity \cite{Bassett2011,Bassett2013,Bassett2014,Bassett2015,Heitger2012,Mantzaris2013} and task-evoked activity \cite{Megumi2015,Floyer2006,Ghilardi2000}, but the relationships between the two are not well understood. Using both types of information, we derive a hybrid community structure representing groups of brain regions that can show both similar functional connectivity and similar values of the annotation while healthy adult individuals perform a discrete sequence production task. We use this approach to ask the question of whether the degree to which the annotation is incorporated in the community structure relates to learning. To address this question, we study the relationship between (i) a mutual information between activity and connectivity, and (ii) learning rates estimated either for each of 10 training levels separately, or for all 10 training levels combined. We observed that annotation contribution to community structure correlates negatively with both types of learning rates, while a non-parametric permutation-based null model does not. These results suggest that faster learning rates are accompanied by decreased similarity between functional connectivity and decaying activity magnitude. It is important to recall that the annotations encode the transformation that activation must undergo from naive to experienced learner -- in other words, the change in activity relative to baseline. Thus, our results indicate that learning occurs best when the regions that show the greatest decrease in activation with learning are not the same as the regions that show dense functional connectivity. Indeed, an autonomy between these two distinct phenotypes of brain dynamics is conducive to better learning.

\section{Conclusion}

A holistic view of the brain must be achieved in order to fully understand the underlying network architecture and dynamics. Here, we propose a method in which we unify functional connectivity with regional activity magnitude to investigate community structure. This method is generalizable to other problems that have the form of an underlying weighted network with superimposed annotation. Here, we apply this method to uncover how the interaction between functional connectivity and regional activity influences learning rate in a motor-visual learning task. Our findings suggest that this relative autonomy between activity and functional connectivity accompanies better learning performance. Our method and results motivate further study of the relationship between activity and connectivity in the performance of other cognitive tasks. Moreover, the method that we develop can also be applied more generally to any scientific question in which one wishes to better understand the relationships between a pattern of connectivity (functional, structural, morphometric), and a characteristic of a brain region (PET, BOLD magnitude, MEG or EEG power, etc.). The method is also amenable to similar questions at smaller spatial scales \cite{betzel2016multiscale}, such as at the level of calcium imaging or microscale connectomics, and to non-neural application areas such as in social \cite{fortunato2014adding}, technological \cite{onnela2012taxonomies}, physical \cite{papadopoulos2016evolution,bassett2015extraction}, and other biological systems \cite{norton2016detecting,henzler2013staged,conaco2012functionalization}.

\section*{Acknowledgments}
This work was supported by the Army Research Office through contract number W911NF-14-1-0679. DSB, SG, ANK, and ACM acknowledge additional support from the John D. and Catherine T. MacArthur Foundation, the Alfred P. Sloan Foundation, the Army Research Laboratory through contract number W911NF-10-2-0022, the National Institute of Mental Health (2-R01-DC-009209-11), the Na-
tional Institute of Child Health and Human Development (1R01HD086888-01), the Office of Naval Research, and the National Science Foundation (BCS-1441502, NSF PHY-1554488, and BCS-1631550). The content is solely the responsibility of the authors and does not necessarily represent the official views of any of the funding agencies.

\section*{References}

\bibliographystyle{elsarticle-harv}
\bibliography{andrew_v4bib}

\end{document}